\documentclass[superscriptaddress, onecolumn, amsmath, amssymb, aps, prx, longbibliography]{revtex4-2}

\usepackage{soul,xcolor}
\usepackage{babel}
\usepackage{natbib}
\usepackage{graphicx}
\usepackage{amsmath}
\usepackage{dcolumn}
\usepackage{bm} 
\usepackage{float}
\usepackage[colorlinks,
            urlcolor=blue,
            linkcolor=blue,
           anchorcolor=blue,
            citecolor=blue]{hyperref}
\usepackage{mfirstuc}

\RequirePackage[explicit]{titlesec}
\renewcommand{\thesubsection}{\Alph{subsection}}
\titleformat{\section}
  {\large\rmfamily\bfseries}
  {\thesection.}
  {0em}
  {\capitalisewords{#1}}
  []
\titleformat{name=\section,numberless}
  {\large\rmfamily\bfseries}
  {}
  {0em}
  {\capitalisewords{#1}}
  [] 
\titleformat{\subsection}[runin]
  {\rmfamily\bfseries}
  {\thesubsection.}
  {0em}
  {#1}
  []
  
\titlespacing*{\section}{0pt}{4ex plus .0ex minus .0ex}{1ex plus .0ex}
\titlespacing*{\subsection}{0pt}{2ex plus .0ex minus .0ex}{.0ex plus .0ex}

\begin{document}
\setstcolor{red}

\title{\textcolor{black}{Riemann-Silberstein geometric phase in 4D polarization space}}

\author{Yuqiong Cheng}\thanks{These authors contributed equally.}
\affiliation{Department of Physics, City University of Hong Kong, Kowloon, Hong Kong, China}
\author{Yuan-Song Zeng}\thanks{These authors contributed equally.}
\affiliation{Department of Electrical Engineering, City University of Hong Kong, Kowloon, Hong Kong, China}
\affiliation{State Key Laboratory of Terahertz and Millimeter Waves, City University of Hong Kong, Kowloon, Hong Kong, China}
\author{Wanyue Xiao}
\affiliation{Department of Physics, City University of Hong Kong, Kowloon, Hong Kong, China}
\author{Tong Fu}
\affiliation{Department of Physics, City University of Hong Kong, Kowloon, Hong Kong, China}
\author{Jiajun Wu}
\affiliation{Department of Electrical Engineering, City University of Hong Kong, Kowloon, Hong Kong, China}
\author{Geng-Bo Wu}\email{bogwu2@cityu.edu.hk}
\affiliation{Department of Electrical Engineering, City University of Hong Kong, Kowloon, Hong Kong, China}
\affiliation{State Key Laboratory of Terahertz and Millimeter Waves, City University of Hong Kong, Kowloon, Hong Kong, China}
\author{Din Ping Tsai}\email{dptsai@cityu.edu.hk}
\affiliation{Department of Physics, City University of Hong Kong, Kowloon, Hong Kong, China}
\affiliation{Department of Electrical Engineering, City University of Hong Kong, Kowloon, Hong Kong, China}
\affiliation{State Key Laboratory of Terahertz and Millimeter Waves, City University of Hong Kong, Kowloon, Hong Kong, China}
\author{Shubo Wang}\email{shubwang@cityu.edu.hk}
\affiliation{Department of Physics, City University of Hong Kong, Kowloon, Hong Kong, China}
\date{\today}
            
\begin{abstract}
\noindent\textcolor{black}{Geometric phase is a far-reaching concept in quantum and classical physics. The first discovered geometric phase, the Pancharatnam-Berry (PB) phase, has profoundly shaped nanophotonics through metasurfaces. However, the PB phase arises from SU(2) polarization evolution and is constrained to a 2D polarization space, failing to capture the full polarization degrees of freedom. We generalize geometric phase to the 4D Riemann-Silberstein (RS) space that simultaneously describes electric, magnetic, and hybrid electric-magnetic polarizations. We show that SU(4) polarization evolution can generate a new geometric phase, the RS phase, alongside the PB phase. Unlike the PB phase that typically manifests in circularly polarized light, the RS phase can emerge in arbitrarily polarized light. Together, they enable a high-dimensional geometric framework for light propagation across general interfaces. We reveal that the phase shifts governed by Fresnel equations are direct manifestations of the RS-space geometric phases, integrating a century-old wave theory into this paradigm. We experimentally validate the framework using metasurfaces and achieve high-dimensional wavefront manipulation. Our work offers fundamental insights into the geometric nature of light-matter interactions, with implications for topological and non-Abelian physics in classical wave systems.}
\end{abstract}

\maketitle


\section*{Introduction}
\noindent
Geometric phases emerge from state evolution in parameter space and provide a unified framework for understanding diverse phenomena in quantum and classical physics {\cite{RN54,RN53,shapere1989geometric,RN51}}. These phases have been extensively studied in various physical systems, including quantum particles \cite{RN53,RN49,RN58,RN48}, condensed matter \cite{RN47,RN46}, and classical wave systems \cite{RN45,chiao1986manifestations,RN43,RN42,RN41}. In optics, geometric phases can give rise to intriguing phenomena such as spin-orbit interactions \cite{RN40} and photonic topological states \cite{RN39,RN38,RN37,RN36}, providing important insights into the geometric and topological properties of optical fields \cite{RN35,RN34,RN33} and enabling novel mechanisms for light manipulation \cite{RN32,RN31,RN30,RN29,kim2022metasurface,xiong2023breaking}. 

The PB phase has recently attracted significant attention for its crucial role in \textcolor{black}{topological meta-optics and structured light manipulation} \cite{RN27,RN26,RN25,RN24,RN23,song2021plasmonic,RN22,RN21,zeng2025metalasers}. This phase arises from SU(2) \textcolor{black}{evolution of electric} polarization on the Poincaré sphere \cite{RN20,RN19}, which is a 2D space describing the polarization of a two-component \textcolor{black}{spinor}. However, light is an electromagnetic wave comprising an electric field $\mathbf{E}$ and a magnetic field $\mathbf{H}$, \textcolor{black}{which can be described as a bispinor in the Dirac-like formulations of Maxwell's equations} \cite{barnett2014optical,alpeggiani2018}. For monochromatic waves, both $\mathbf{E}$ and $\mathbf{H}$ are two-component vector fields within the local frame defined by their polarization ellipses \cite{RN35,RN18}, and their polarizations can be different. Consequently, the complete polarization state of a monochromatic electromagnetic wave resides in a 4D Hilbert space (i.e., direct sum of electric and magnetic polarization spaces), termed RS space in this paper, which simultaneously characterizes electric, magnetic, and hybrid electric-magnetic polarizations. The SU(4) evolution of the 4D complete polarization can give rise to nontrivial geometric phases beyond the conventional PB phase, which have thus far remained elusive. \textcolor{black}{These RS-space geometric phases are essential to establishing a unified geometric framework for light-matter interactions involving complex materials or structured light.} 

\begin{figure*}[htp]
\centering
\includegraphics{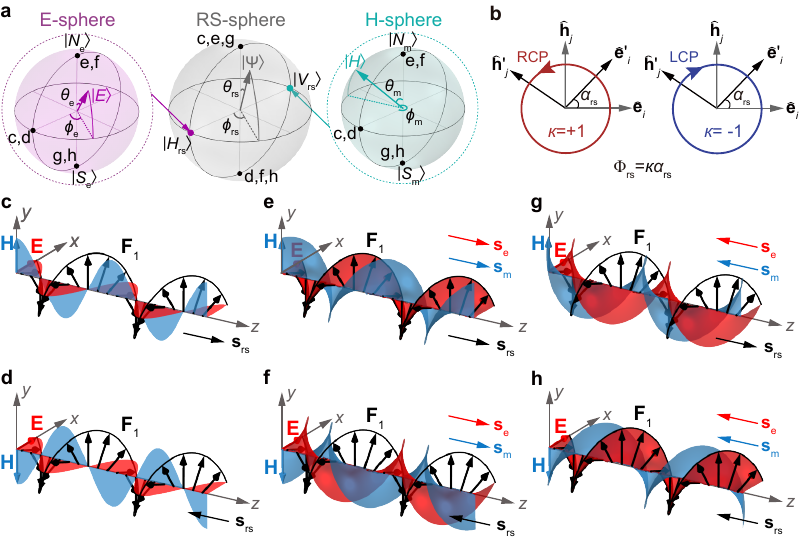}
\caption{\textbf{Complete 4D polarization of electromagnetic fields}. (a) Poincaré hypersphere representation of 4D electromagnetic polarization, where the points “c-h” represent the polarizations of the plane waves in the panels (c-h), respectively. (b) RS geometric phase induced by the rotation of the local constitutive frame. Electric field (in red), magnetic field (in blue), and hybrid RS field (in black) of (c) $+z$-propagating linearly polarized plane wave, (d) $-z$-propagating linearly polarized plane wave, (e) $+z$-propagating RCP plane wave, (f) $-z$-propagating RCP plane wave, (g) $+z$-propagating LCP plane wave, and (h) $-z$-propagating LCP plane wave. The electric spin density \textcolor{black}{$\mathbf{s_\mathrm{e}}$}, magnetic spin density \textcolor{black}{$\mathbf{s_\mathrm{m}}$}, and RS spin density \textcolor{black}{$\mathbf{s_\mathrm{rs}}$} are denoted by the red, blue, and black arrows, respectively.}
\label{fig1}
\end{figure*}  

In this work, \textcolor{black}{we expand the geometric phase paradigm from the 2D polarization space to the 4D polarization space.} We identify a new class of geometric phase, the RS phase, which \textcolor{black}{can emerge in general electromagnetic waves and} is governed by the hybrid electric-magnetic polarization, in contrast to the conventional PB phase \textcolor{black}{associated with individual electric (magnetic) polarization}. \textcolor{black}{The RS and PB phases complement each other, and together enable a high-dimensional geometric framework for light propagation across general interfaces. This framework reveals a hidden geometric nature of the phase shifts due to light transmission and reflection at interfaces, a classical phenomenon governed by Fresnel equations. We experimentally demonstrate the RS phase and verify the framework at microwave frequencies using metasurfaces, which gives rise to high-dimensional wavefront deflection that cannot \textcolor{black}{be achieved} in conventional metasurface systems. The results provide new insights into light-matter interactions at interfaces and expand the toolkit for exploring geometric and topological properties of light.} 

\subsection*{\large{\textcolor{black}{Geometric phases in RS space}}}\

\noindent \textcolor{black}{The polarization state of light is conventionally described as a two-component spinor governed by an effective Hamiltonian $\hat{\mathcal{H}}=\hat{\rho}-\frac{1}{2} \hat{\text{I}}_2=\frac{1}{2} \mathbf{S} \cdot \hat{\boldsymbol{\sigma}}$, where $\hat{\rho}=|E\rangle\langle E|$ is the polarization projector, $\hat{\text{I}}_2$ is the $2 \times 2$ identity matrix, $\hat{\boldsymbol{\sigma}}$ is the vector of Pauli spin matrices, and $\mathbf{S}$ is the Stokes vector \cite{RN20}. The 2D polarization state $|E\rangle$ can be represented on the Poincaré sphere \cite{RN54}. In our framework, a monochromatic electromagnetic field is described by the wavefuntion $\boldsymbol{\Psi}=\mathbf{E}+\mathrm{i}\mathbf{H}$ (Gaussian units are used throughout the paper), as in the Dirac-like formulation of electromagnetism \cite{bialynicki1996v}. Here,} $\mathbf{E}(\mathbf{r}, t)=E_i(\mathbf{r}, t) \hat{\mathbf{e}}_i+E_j(\mathbf{r}, t) \hat{\mathbf{e}}_j$ is the complex electric field in the local frame with bases $\left(\hat{\mathbf{e}}_i, \hat{\mathbf{e}}_j\right)$, and $\mathbf{H}(\mathbf{r}, t)=H_i(\mathbf{r}, t) \hat{\mathbf{h}}_i+H_j(\mathbf{r}, t) \hat{\mathbf{h}}_j$ is the complex magnetic field in the local frame with bases $(\hat{\mathbf{h}}_i, \hat{\mathbf{h}}_j)$. \textcolor{black}{The complete polarization state $|\Psi\rangle$ is a four-component bispinor governed by the effective Hamiltonian $\hat{\mathcal{H}}=\hat{\rho}-\frac{1}{4} \hat{\text{I}}_4=\frac{1}{2}\mathbf{S}\cdot \hat{\boldsymbol{\lambda}}$, where $\hat{\rho}=|\Psi\rangle\langle\Psi|$ and $\hat{\boldsymbol{\lambda}}$ is the generalized Gell-Mann matrices (Supplementary Note I). This 4D polarization state} exhibits both internal polarizations (i.e., the polarizations of individual $\mathbf{E}$ and $\mathbf{H}$ fields) and external polarization (i.e., the hybrid $\mathbf{E}$-$\mathbf{H}$ polarization). \textcolor{black}{It resides in a 4D RS Hilbert space and can be represented on a Poincaré hypersphere} {\cite{RN17,RN16}, as illustrated in Fig. \ref{fig1}\textbf{a}, which comprises three nested Poincaré spheres: E-sphere characterizing the $\mathbf{E}$ field polarization, H-sphere characterizing the $\mathbf{H}$ field polarization, and RS-sphere characterizing the hybrid $\mathbf{E}$-$\mathbf{H}$ polarization. The E-sphere has the north pole state $\left|N_{\mathrm{e}}\right\rangle$ corresponding to right-handed circularly-polarized (RCP) electric field and the south pole state $\left|S_{\mathrm{e}}\right\rangle$ corresponding to left-handed circularly-polarized (LCP) electric field; The H-sphere has the north pole state $\left|N_{\mathrm{m}}\right\rangle$ corresponding to RCP magnetic field and the south pole state $\left|S_{\mathrm{m}}\right\rangle$ corresponding to the LCP magnetic field. Any other point on the E-sphere or H-sphere denotes a $\mathrm{SU}(2)$ superposition state parameterized by the polar angle $\theta_{\mathrm{e}, \mathrm{m}}$ and azimuthal angle $\phi_{\mathrm{e}, \mathrm{m}}$. Notably, the RS-sphere characterizes the relationship between E-sphere and H-sphere with $\theta_{\mathrm{rs}}$ and $\phi_{\mathrm{rs}}$; its horizontal basis state $\left|H_{\mathrm{rs}}\right\rangle$ corresponds to an arbitrary state on the E-sphere; its vertical basis state $\left|V_{\mathrm{rs}}\right\rangle$ corresponds to an arbitrary state on the H-sphere. Any other point on the RS-sphere denotes a $\operatorname{SU}(4)$ superposition state $|\Psi\rangle$ \textcolor{black}{(see Methods)}. 
\textcolor{black}{The SU(4) evolution of $|\Psi\rangle$ will generally trace out trajectories on all the three spheres, and each sphere with an effective magnetic monopole in the center will contribute to the total geometric phase of $\boldsymbol{\Psi}$ with a weighting. In particular, the polarization evolution on the RS-sphere will generate a new type of geometric phase, termed RS phase, which will be elaborated with concrete examples below.}

In the following, we will focus on paraxial waves involving transverse electric and magnetic fields that are mutually perpendicular. In such cases, $\mathbf{E}$ and $\mathbf{H}$ exhibit the same polarization, and $\boldsymbol{\Psi}$ can be decomposed into a pair of orthogonal RS vectors: $\boldsymbol{\Psi}=\mathbf{F}_1+\mathbf{F}_2$, where $\mathbf{F}_1(\mathbf{r}, t)= E_i(\mathbf{r}, t) \hat{\mathbf{e}}_i+\mathrm{i} H_j(\mathbf{r}, t) \hat{\mathbf{h}}_j$ and $\mathbf{F}_2(\mathbf{r}, t)=E_j(\mathbf{r}, t) \hat{\mathbf{e}}_j+\mathrm{i} H_i(\mathbf{r}, t) \hat{\mathbf{h}}_i$. Consequently, the hybrid $\mathbf{E}$-$\mathbf{H}$ polarization reduces to the polarizations of $\mathbf{F}_1$ and $\mathbf{F}_2$. We note that the RS vectors $\mathbf{F}_1$ and $\mathbf{F}_2$ are composed of \textit{complex} electric and magnetic fields, which are different from the conventional RS vector comprising \textit{real} electric and magnetic fields \cite{RN15,RN13,RN57,RN12,RN11}.

We use electromagnetic plane waves as an example to illustrate the 4D polarization and its representation on the Poincaré hypersphere. We first consider a plane wave with linearly polarized electric and magnetic fields: $\boldsymbol{\Psi}=(\hat{\mathbf{e}}_x+\mathrm{i}\kappa\hat{\mathbf{h}}_y) E e^{\mathrm{i} \kappa k_0 z-\mathrm{i} \omega t}$, where $\kappa= \pm 1$ denotes the sign of the propagation direction relative to $+z$ axis. Figures \ref{fig1}\textbf{c} and \ref{fig1}\textbf{d} show the instantaneous $\mathbf{E}$, $\mathbf{H}$, and $\mathbf{F}_1$ fields for $\kappa=1$ and $\kappa= -1$, respectively. The polarizations of $\mathbf{E}$ and $\mathbf{H}$ fields are represented by points “c” and “d” on the E-sphere and H-sphere in Fig. \ref{fig1}\textbf{a}. The RS field $\mathbf{F}_1$ is RCP for $\kappa= 1$ and LCP for $\kappa= -1$, which are represented by points “c” and “d” on the RS-sphere in Fig. \ref{fig1}\textbf{a}. Naturally, we can introduce an RS spin density $\textcolor{black}{\mathbf{s}}_{\mathrm{rs}}=\frac{1}{16 \pi \omega} \operatorname{Im}\left[\mathbf{F}_1^* \times \mathbf{F}_1\right]=\frac{1}{\omega c} \mathbf{P}$, where $\mathbf{P}=\frac{c}{8 \pi} \operatorname{Re}\left[\mathbf{E}^* \times \mathbf{H}\right]$ is the time-averaged Poynting vector. This RS spin \textcolor{black}{originates from the intrinsic chirality of RS fields with the handedness defined by $\kappa= \pm 1$}.

We further consider a plane wave with circularly polarized electric and magnetic fields: $\boldsymbol{\Psi}= [\left(\hat{\mathbf{e}}_x+\mathrm{i}\sigma_{\mathrm{e}}\hat{\mathbf{e}}_y\right)+\mathrm{i}\kappa(\hat{\mathbf{h}}_y-\mathrm{i}\sigma_{\mathrm{m}}\hat{\mathbf{h}}_x)] E e^{\mathrm{i} \kappa k_0 z-\mathrm{i} \omega t}$, where $\sigma_{\mathrm{e}}= \pm 1$ denotes the electric spin, $\sigma_{\mathrm{m}}= \pm 1$ denotes the magnetic spin, and $\kappa= \pm 1$ denotes the RS spin. Note that $\sigma_{\mathrm{e}}=\sigma_{\mathrm{m}}$ for plane waves. The total spin density can be expressed as $\textcolor{black}{\mathbf{s}}=\frac{1}{16 \pi \omega} \operatorname{Im}\left[\boldsymbol{\Psi}^* \times \boldsymbol{\Psi}\right]=\textcolor{black}{\mathbf{s}}_{\mathrm{e}}+\textcolor{black}{\mathbf{s}}_{\mathrm{m}}+\textcolor{black}{\mathbf{s}}_{\mathrm{rs}}$, where $\textcolor{black}{\mathbf{s}}_{\mathrm{e}}=\frac{1}{16 \pi \omega} \operatorname{Im}\left[\mathbf{E}^* \times \mathbf{E}\right]$ is the electric spin density, $\textcolor{black}{\mathbf{s}}_{\mathrm{m}}=\frac{1}{16 \pi \omega} \operatorname{Im}\left[\mathbf{H}^* \times \mathbf{H}\right]$ is the magnetic spin density, and $\textcolor{black}{\mathbf{s}}_{\mathrm{rs}}=\frac{1}{16 \pi \omega} \operatorname{Im}\left[\mathbf{F}_1^* \times \mathbf{F}_1\right]+ \frac{1}{16 \pi \omega} \operatorname{Im}\left[\mathbf{F}_2^* \times \mathbf{F}_2\right]=\frac{1}{\omega c} \mathbf{P}$ is the RS spin density. Notably, the sum of $\textcolor{black}{\mathbf{s}}_{\mathrm{e}}$ and $\textcolor{black}{\mathbf{s}}_{\mathrm{m}}$ corresponds to the conventional optical spin density that has been extensively studied in recent years \cite{RN40,RN10,RN8,RN7,RN6,RN5,RN4}, while the RS spin density $\textcolor{black}{\mathbf{s}}_{\mathrm{rs}}$ has been largely overlooked \cite{RN11,golat2025electromagnetic}. Figure \ref{fig1}\textbf{e-h} shows the instantaneous $\mathbf{E}$, $\mathbf{H}$, $\mathbf{F}_1$, and spin directions for the plane waves with $\left(\sigma_{\mathrm{e}}, \sigma_{\mathrm{m}}, \kappa\right)= (+1,+1,+1),(+1,+1,-1),(-1,-1,+1)$, and $(-1,-1,-1)$, respectively. Their 4D polarizations are represented on the Poincaré hypersphere in Fig. \ref{fig1}\textbf{a} as the points “e”, “f”, “g”, and “h”, respectively.

The evolution of 4D polarization state $|\Psi\rangle$ simultaneously traces out paths on the E-sphere, H-sphere, and RS-sphere. The electric (magnetic) polarization evolution can generate the PB phase $\Phi_{\mathrm{pb}}$, which is proportional to the solid angle subtended by the area enclosed by the evolution path on E-sphere (H-sphere) \cite{RN54}. Notably, the electric and magnetic polarization evolutions are intrinsically linked through the Maxwell equations. For the considered paraxial waves, the electric and magnetic polarization evolutions trace the same path and contribute to the same PB phase. \textcolor{black}{Additionally, the polarization evolution of $\mathbf{F}_{1,2}$ will generate the RS phase  $\Phi_{\mathrm{rs}}$,} which is proportional to the solid angle subtended by the area enclosed by the evolution path on the RS-sphere (Supplementary Note III). Importantly, $\Phi_{\mathrm{rs}}$ is independent of $\Phi_{\mathrm{pb}}$ because $\Phi_{\mathrm{pb}}$ manifests in the basis states $\left|H_{\mathrm{rs}}\right\rangle$ and $\left|V_{\mathrm{rs}}\right\rangle$. Therefore, the total geometric phase due to 4D polarization evolution is $\Phi_{\mathrm{tot}}=\Phi_{\mathrm{pb}}+\Phi_{\mathrm{rs}}$.

The PB phase can be attributed to the coupling between electric (magnetic) spin and rotation of local coordinate frame, where the frame rotation induces a phase variation of the circularly polarized electric (magnetic) field \cite{RN40}. Similarly, the RS phase can be attributed to the coupling between RS spin and rotation of local \textit{constitutive} frame $(\hat{\mathbf{e}}_i, \hat{\mathbf{h}}_j)[\text{or } (\hat{\mathbf{e}}_j, \hat{\mathbf{h}}_i)]$, as shown in Fig. \ref{fig1}\textbf{b}. A rotation of the constitutive frame by angle $\alpha_{\mathrm{rs}}$ leads to the transformation of the circularly polarized RS vector: $\mathbf{F}_{1,2} \rightarrow \mathrm{e}^{\mathrm{i} \kappa \alpha_{\mathrm{rs}}} \mathbf{F}_{1,2}$. The phase $\kappa \alpha_{\mathrm{rs}}$ corresponds to the RS geometric phase, which is proportional to the RS spin $\kappa$ and the rotation angle $\alpha_{\mathrm{rs}}$ of the local constitutive frame.

\begin{figure*}[tp]
\centering
\includegraphics[width=0.85\linewidth]{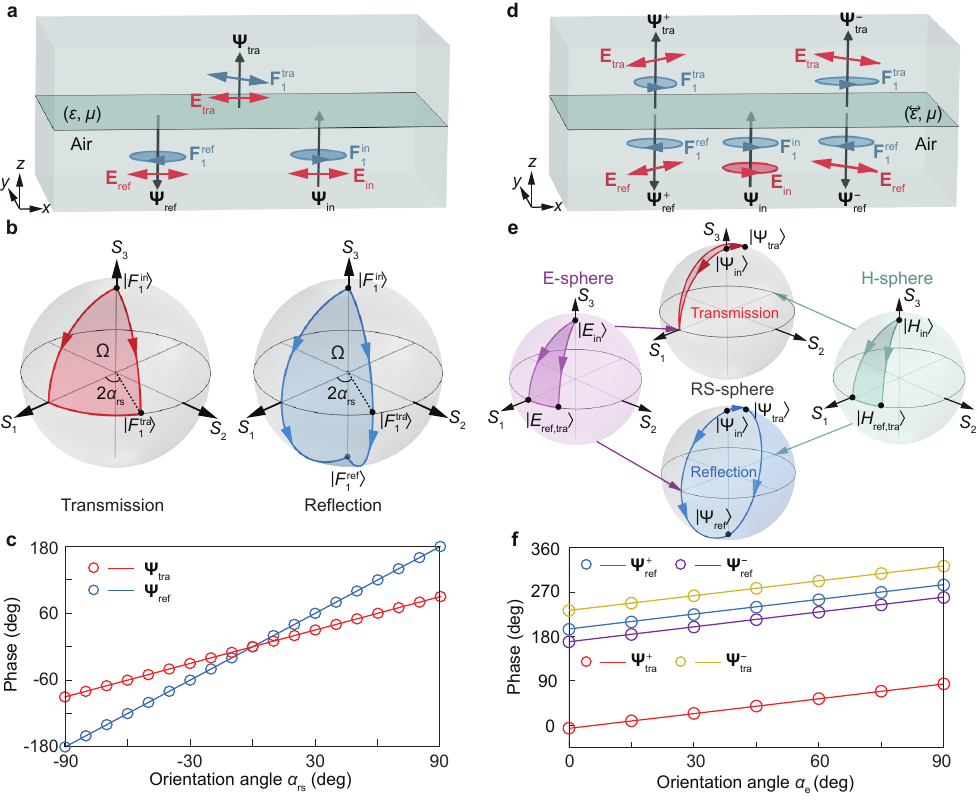}
\caption{\textbf{RS geometric phase at  interfaces}. (a) A plane wave with linearly polarized electric and magnetic fields impinges on an interface of isotropic media, where the RS polarization (blue arrows) undergoes evolutions but electric polarization (red arrows) remains unchanged. (b) Representation of the RS polarization evolution on the RS-sphere for the wave transmission and reflection. We choose the point on ${S}_1$ axis as the reference polarization. (c) Transmission and reflection phases for different orientation angles of the transmitted RS polarization. The circles denote the RS geometric phase. (d) A plane wave with circularly polarized electric and magnetic fields impinges on an interface of anisotropic media, where both the electric polarization (red arrows) and the RS polarization (blue arrows) undergo evolutions. (e) Representation of the 4D polarization evolution on the Poincaré hypersphere for the wave transmission and reflection. (f) Transmission and reflection phases for different orientation angle of the electric polarization. The circles denote the total geometric phase.}
\label{fig2}
\end{figure*}  

\subsection*{\large{\textcolor{black}{RS phase at interfaces}}}\ 

\noindent\textcolor{black}{The RS phase can emerge in light transmission and reflection at an interface due to 4D polarization evolution, which is determined by the eigen polarization states of the two media forming the interface. Remarkably, the RS phase reveals the hidden geometric nature of the phase shifts governed by Fresnel equations, a cornerstone of classical optics.} As shown in Fig. \ref{fig2}\textbf{a}, we consider that a plane wave propagates along $+z$ direction and normally impinges on the surface of an isotropic medium with permittivity $\varepsilon$ and permeability $\mu$, \textcolor{black}{giving rise to RS polarization evolution}. The incident electric (magnetic) field is linearly polarized in $x$ ($y$) direction. We denote the RS polarization states of the incident, reflected, and transmitted waves as $\left|F_1^{\text {in}}\right\rangle,\left|F_1^{\text {ref}}\right\rangle$, and $\left|F_1^{\text {tra}}\right\rangle$, respectively. The transmission involves the polarization evolution $\left|F_1^{\text {in}}\right\rangle \rightarrow\left|F_1^{\text {tra}}\right\rangle$, while the reflection involves the polarization evolution $\left|F_1^{\text {in}}\right\rangle \rightarrow\left|F_1^{\text {tra}}\right\rangle \rightarrow\left|F_1^{\text {ref}}\right\rangle$. These RS polarization evolutions induce the RS phases $\Phi_{\mathrm{rs}}^{\text {tra}}$ and $\Phi_{\mathrm{rs}}^{\mathrm{ref}}$, which manifest in the transmitted and reflected waves, respectively. As an example, we assume the medium is a lossless metal. In this case, $\left|F_1^{\mathrm{in}}\right\rangle=\frac{1}{\sqrt{2}}(\hat{\mathbf{e}}_x+\mathrm{i} \hat{\mathbf{h}}_y),\left|F_1^{\mathrm{ref}}\right\rangle=\frac{1}{\sqrt{2}}(\hat{\mathbf{e}}_x-\mathrm{i} \hat{\mathbf{h}}_y)$, and \textcolor{black}{$\left|F_1^{\text {tra}}\right\rangle=\frac{\sqrt{\left|{\mu}\right|}}{\sqrt{\left|{\mu}\right|+\left|{\varepsilon}\right|}}(\hat{\mathbf{e}}_x+\frac{\mathrm{i\sqrt{\varepsilon}}}{\sqrt{\mu}}\hat{\mathbf{h}}_y)$}. Notably, $\left|F_1^{\text {tra}}\right\rangle$ is linearly polarized in the constitutive frame with orientation angle \textcolor{black}{$\alpha_{\mathrm{rs}}=\text{arctan}(\frac{\mathrm{i\sqrt{\varepsilon}}} {\sqrt{\mu}})$}, which locates on the equator of the RS-sphere with azimuthal angle $\phi_{\mathrm{rs}}=2 \alpha_{\mathrm{rs}}$. Thus, changing the material properties $(\varepsilon,\mu)$ leads to rotation of local constitutive frame and thus the variation of $\phi_{\mathrm{rs}}$. Figure \ref{fig2}\textbf{b} shows the polarization evolution paths on the RS-sphere for the transmission and reflection. We calculate the RS phases $\Phi_{\mathrm{rs}}^{\text {tra}}$ and $\Phi_{\mathrm{rs}}^{\text {ref}}$ by evaluating the solid angles $\Omega$ enclosed by the paths. The results are shown as the circles in Fig. \ref{fig2}\textbf{c} for different orientation angle $\alpha_{\mathrm{rs}}$, which agree with the transmission and reflection phases (solid lines) predicted by the Fresnel equations \cite{RN1}. This relationship holds for normal incidence at general interfaces between isotropic media (Supplementary Note IV).

\begin{figure*}[htp]
\centering
\includegraphics[width=0.9\linewidth]{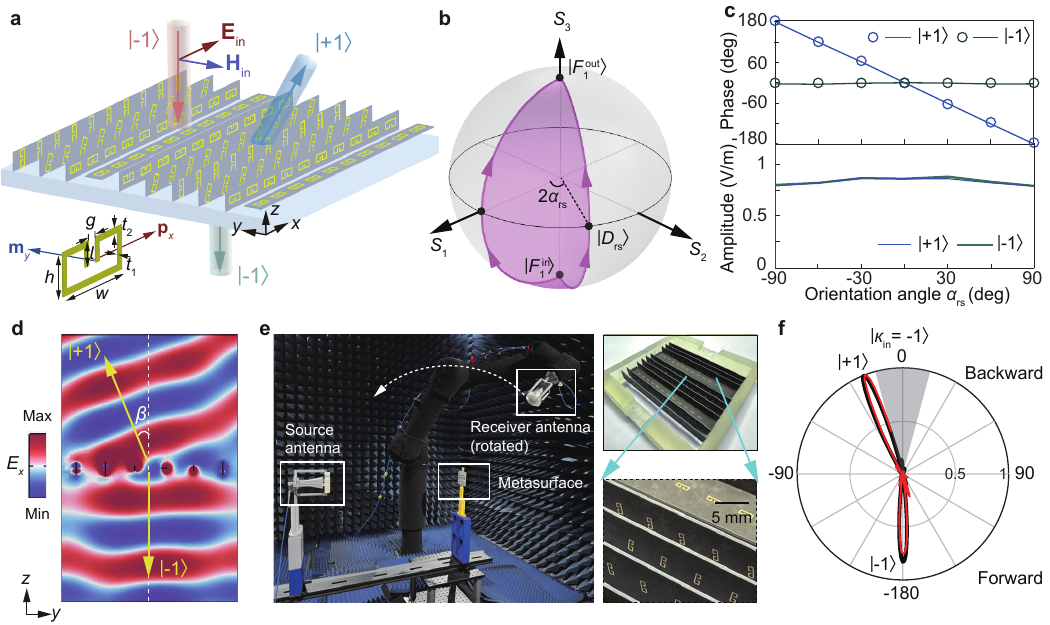}
\caption{\textbf{\textcolor{black}{RS geometric phase at a metasurface}}. (a) Schematic of the RS metasurface under the normally incident plane wave with linearly polarized electric field. The inset shows the structure of the meta-atom. (b) RS polarization evolution on the RS-sphere induced by the metasurface.
(c) Theoretical RS geometric phase (symbols), simulated phase (solid lines), and simulated amplitude for the forward and backward scattering fields as a function of the orientation angle $\alpha_\mathrm{rs}$ of the RS dipole. (d) Simulated electric field scattered by the RS metasurface. (e) Experiment setup and the fabricated metasurface prototype. (f) Simulated and experimentally measured far-field intensity pattern. The shaded area marks the measurement blind zone [-15 deg, 15 deg] due to the source antenna obstruction.}
\label{fig3}
\end{figure*}  

Both the PB and RS phases can emerge at an interface involving anisotropic media. We consider the interface between air and a non-magnetic medium that has anisotropic in-plane permittivity $\overleftrightarrow{\varepsilon}$ with elements $\varepsilon_{xx}=\varepsilon+\delta\cos\phi, \varepsilon_{yy}=\varepsilon-\delta\cos\phi, \varepsilon_{xy}=\delta \sin \phi e^{-\mathrm{i} \theta}$, and $\varepsilon_{yx}=\delta \sin \phi e^{\mathrm{i} \theta}$. The medium supports two orthogonal 4D eigen polarization states.  As shown in Fig. \ref{fig2}\textbf{d}, under the normal incidence of a plane wave with polarization $\left|\Psi_{\text {in}}\right\rangle=\frac{1}{2}[(\hat{\mathbf{e}}_x+\mathrm{i} \hat{\mathbf{e}}_y)+\mathrm{i}(\hat{\mathbf{h}}_y-\mathrm{i} \hat{\mathbf{h}}_x)]$, two transmitted waves $\left|\Psi_{\text {tra}}^{ \pm}\right\rangle=\frac{1}{\sqrt{1+|\varepsilon \pm \delta|}}(\mathbf{E}_{ \pm}+\mathrm{i} \sqrt{\varepsilon \pm \delta} \mathbf{H}_{ \pm})$ and two reflected waves $\left|\Psi_{\text {ref}}^{ \pm}\right\rangle=\frac{1}{\sqrt{2}}\left(\mathbf{E}_{ \pm}-\mathrm{i} \mathbf{H}_{ \pm}\right)$ emerge, where $\mathbf{E}_{+}=\cos \frac{\phi}{2} \hat{\mathbf{e}}_x+ \sin \frac{\phi}{2} e^{\mathrm{i} \theta} \hat{\mathbf{e}}_y, \mathbf{E}_{-}=\sin \frac{\phi}{2} e^{-\mathrm{i} \theta} \hat{\mathbf{e}}_x-\cos \frac{\phi}{2} \hat{\mathbf{e}}_y, \mathbf{H}_{+}=\cos \frac{\phi}{2} \hat{\mathbf{h}}_y-\sin \frac{\phi}{2} e^{\mathrm{i} \theta} \hat{\mathbf{h}}_x$, and $\mathbf{H}_{-}=\sin \frac{\phi}{2} e^{-\mathrm{i} \theta} \hat{\mathbf{h}}_y+\cos \frac{\phi}{2} \hat{\mathbf{h}}_x$. Correspondingly, four polarization evolutions occur at the interface: $\left|\Psi_{\text {in}}\right\rangle \rightarrow\left|\Psi_{\text {tra}}^{+}\right\rangle, \left|\Psi_{\text {in}}\right\rangle \rightarrow\left|\Psi_{\text {tra}}^{-}\right\rangle, \left|\Psi_{\text {in}}\right\rangle \rightarrow\left|\Psi_{\text {tra}}^{+}\right\rangle \rightarrow\left|\Psi_{\text {ref}}^{+}\right\rangle$, and $\left|\Psi_{\text {in}}\right\rangle \rightarrow\left|\Psi_{\text {tra}}^{-}\right\rangle \rightarrow\left|\Psi_{\text {ref}}^{-}\right\rangle$. As an example, Fig. \ref{fig2}\textbf{e} depicts the polarization evolutions associated with $\left|\Psi_{\text {tra}}^{+}\right\rangle$ and $\left|\Psi_{\text {ref}}^{+}\right\rangle$ for $\varepsilon=1+\mathrm{i}, \delta=2, \phi=30^{\circ}$, and $\theta=0^{\circ}$. Notably, the electric and magnetic polarizations evolve along the same pathway on the E-sphere and H-sphere, which are independent of the RS polarization evolution on the RS-sphere. Figure \ref{fig2}\textbf{f} shows the total geometric phases (i.e., sum of PB and RS phases) induced by the four polarization evolutions for different $\alpha_{\mathrm{e}}=\frac{\phi}{2}$. The geometric phases (circles) agree with the transmission and reflection phases (solid lines) predicted by the Fresnel equations. Thus, the well-known reflection and transmission phases at interfaces can be interpretated as the geometric phases induced by 4D polarization evolution, \textcolor{black}{uncovering the geometric origin of this fundamental optical phenomena.}

\subsection*{\large{\textcolor{black}{RS phase at metasurfaces}}}\

\noindent The RS phase can also emerge at artificial interfaces such as electromagnetic metasurfaces, due to the modulation of local electric and magnetic responses. To demonstrate this, we propose an RS metasurface comprising metallic split rings with the same geometric dimensions, as shown in Fig. \ref{fig3}\textbf{a}. The metasurface is under the normal incidence of a plane wave with linearly polarized electric and magnetic fields. The corresponding incident RS field is $\mathbf{F}_1^{\mathrm{in}}=(\hat{\mathbf{e}}_x-\mathrm{i} \hat{\mathbf{h}}_y) E e^{-\mathrm{i} k_0 z-\mathrm{i} \omega t}$; its polarization state can be labelled by the RS spin as $\left|\kappa_{\mathrm{in}}=-1\right\rangle$. The incident wave excites an electric dipole $p_x$ and a magnetic dipole $m_y$ in the meta-atoms, forming a RS dipole $\mathbf{D}_{\mathrm{rs}}=p_x \hat{\mathbf{e}}_x+ \mathrm{i} {m_y} \hat{\mathbf{h}}_y$ and generating scattering fields in forward and backward directions. The scattering RS field is $\mathbf{F}_1^{\text {out}}=(\hat{\mathbf{e}}_x+\mathrm{i} \kappa_{\text {out}} \hat{\mathbf{h}}_y) E e^{\mathrm{i} \kappa_{\text {out}} k_0 z-\mathrm{i} \omega t}$ with polarization state $\left|\kappa_{\text {out}}= \pm 1\right\rangle$. Denoting the polarization evolution as $\left|\kappa_{\text {in}}\right\rangle \rightarrow\left|\kappa_{\text {out}}\right\rangle$, the RS phase emerges in the cross-polarized channel $|-1\rangle \rightarrow|+1\rangle$ but vanishes in the co-polarized channel $|-1\rangle \rightarrow|-1\rangle$.

\begin{figure*}[htp]
\centering
\includegraphics[width=\linewidth]{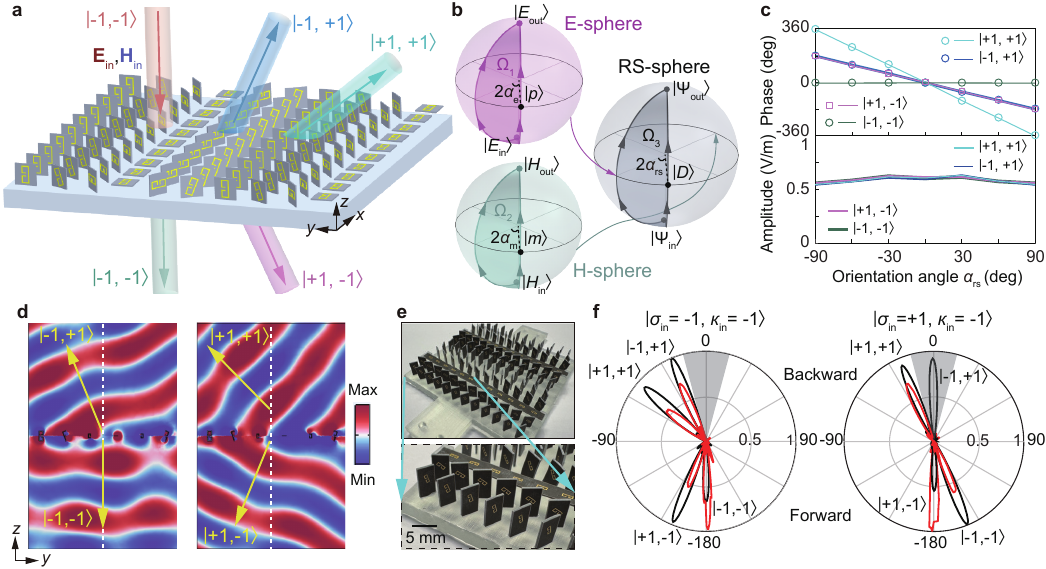}
\caption{\textbf{\textcolor{black}{Complementary RS and PB phases at a metasurface}}. (a) Schematic of the quadruplex RS meta-deflector under the normal incidence of a plane wave with LCP electric field. (b) 4D polarization evolution on the Poincaré hypersphere induced by the higher-order RS metasurface. (c) Theoretical geometric phase \textcolor{black}{(symbols)}, simulated phase \textcolor{black}{(solid lines)}, and simulated amplitude in four polarization evolution channels as a function of the orientation angle of RS dipole in the meta-atom. (d) Simulated electric field distribution of four output waves. (e) \textcolor{black}{Fabricated metasurface prototype.} (f) Simulated and experimentally measured far-field intensity pattern under orthogonal incidence. The shaded area marks the measurement blind zone [-15 deg, 15 deg] due to the source antenna obstruction.}
\label{fig4}
\end{figure*} 

Figure \ref{fig3}\textbf{b} shows the RS polarization evolution in the channel $|-1\rangle \rightarrow|+1\rangle$. The incident state $\left|F_1^{\text {in}}\right\rangle$ and backward scattering state $\left|F_1^{\text {out}}\right\rangle$ are represented by the south and north poles, respectively. The polarization state of the RS dipole $\mathbf{D}_{\mathrm{rs}}$, denoted as $\left|D_{\mathrm{rs}}\right\rangle$, is located on the equator. The orientation angle of $\mathbf{D}_{\mathrm{rs}}$ in the constitutive frame is $\alpha_{\mathrm{rs}}=\text{arctan}(\frac{\mathrm{i} m_y}{p_x})$. Rotating the meta-atoms around $x$ or $y$ direction will alter $\alpha_{\mathrm{rs}}$ (Supplementary Note V) and change the azimuthal angle on the RS-sphere. The resulting RS phase is $\Phi_{\mathrm{rs}}=-2 \alpha_{\mathrm{rs}}$. Notably, the PB phase vanishes in this case because the meta-atom rotation around $x$ or $y$ direction does not change the electric or magnetic polarization. Figure \ref{fig3}\textbf{c} shows the phases and amplitudes of the scattering fields as a function of the orientation angle $\alpha_{\mathrm{rs}}$ of the RS dipole $\mathbf{D}_{\mathrm{rs}}$. The simulated phases (solid lines) agree with the RS phase (circles) obtained by evaluating the solid angle in the RS-sphere. Notably, the phase of the backward scattering field $|+1\rangle$ exhibits a linear relationship with $\alpha_{\mathrm{rs}}$, while the phase of the forward scattering field $|-1\rangle$ is independent of $\alpha_{\mathrm{rs}}$. In addition, the scattering amplitudes remain approximately constant for different $\alpha_{\mathrm{rs}}$.

The RS metasurface in Fig. \ref{fig3}\textbf{a} can generate RS phase gradient along $y$ direction, which can deflect the incident wave. Figure \ref{fig3}\textbf{d} shows the simulated electric field $E_x$ scattered by the RS metasurface. We note that only the backward scattering field is deflected into oblique direction, and the wavefront is consistent with the deflection angle $\beta=\arcsin (\frac{1}{k_0} \frac{\Delta \Phi_{\mathrm{rs}}}{\Delta y})$ predicted by the generalized Snell's law \cite{RN59}. The deflection direction can be reversed by flipping the RS spin (i.e., propagation direction) of the incident plane wave (Supplementary Note VI). We conduct microwave experiments to verify the theory by using the experiment setup and metasurface in Fig. \ref{fig3}\textbf{e} (Methods). The measured far-field intensity scattered by the metasurface is shown as the red line in Fig. \ref{fig3}\textbf{f}, which agrees well with the simulation result denoted by the black line. Only the backward scattering lobe exhibits a deflection angle, confirming the validity of the theory.

\begin{figure*}[htp]
\centering
\includegraphics{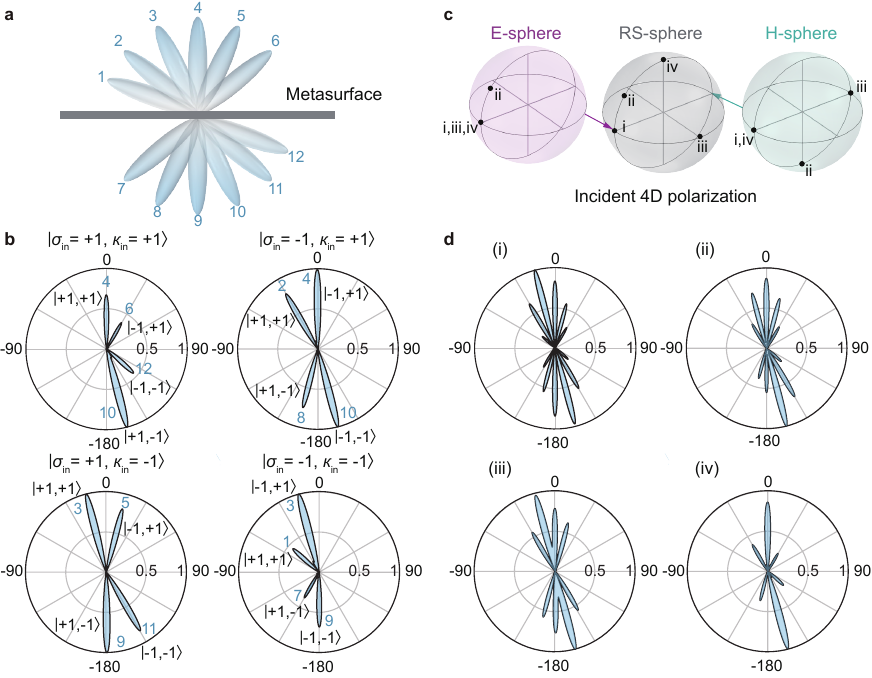}
\caption{\textbf{Reconfigurable wave deflection by the general RS metasurface}. (a) Schematic of the multiplexed beam forming with twelve distinct output wavefronts. (b) Simulated far-field intensity patterns under the normal incidence of plane waves with orthogonal 4D polarizations. The lobes are labelled in accordance with (a). (c) Incident 4D polarizations (labelled as “i-iv") for achieving reconfigurable far-field intensity patterns. (d) Far-field intensity patterns with different number of lobes induced by the incident polarizations “i-iv” in (c).}
\label{fig5}
\end{figure*}

\textcolor{black}{To verify the high-dimensional geometric framework based on the RS and PB phases}, we consider the higher-order RS metasurface in Fig. \ref{fig4}\textbf{a} under the incidence of a plane wave with circularly polarized electric and magnetic fields. Compared to the metasurface in Fig. \ref{fig3}\textbf{a}, \textcolor{black}{the meta-atoms here are further rotated around the local $z$ axis (Methods), which induces electric and magnetic polarization evolutions and gives rise to the additional PB phase $\Phi_{\mathrm{pb}}$.} Consequently, the metasurface generates scattering fields carrying the total geometric phase $\Phi_{\text {tot }}=\Phi_{\mathrm{pb}}+\Phi_{\mathrm{rs}}$. The incident wave field is $\boldsymbol{\Psi}_{\text {in}}=[\left(\hat{\mathbf{e}}_x+\mathrm{i} \sigma_{\text{in}}\hat{\mathbf{e}}_y\right)-\mathrm{i}(\hat{\mathbf{h}}_y-\mathrm{i} \sigma_{\text {in}}\hat{\mathbf{h}}_x)] E e^{-\mathrm{i} k_0 z-\mathrm{i} \omega t}$, which carries the spins $\sigma_{\text {in }}=-1$ and $\kappa_{\text {in }}=-1$. Its electric, magnetic, and RS polarization states are represented by the south poles on the E-sphere, H-sphere, and RS-sphere, respectively, as shown in Fig. \ref{fig4}\textbf{b}. The incident wave excites the 4D dipole in the meta-atoms: $\mathbf{D}=\mathbf{p}+\mathrm{i}{\mathbf{m}}=[(p_x \hat{\mathbf{e}}_x+p_y \hat{\mathbf{e}}_y)+{\mathrm{i}}(m_y \hat{\mathbf{h}}_y+m_x \hat{\mathbf{h}}_x)] e^{-\mathrm{i} \omega t}$. The polarization states of the electric, magnetic, and RS dipoles are represented by $|p\rangle,|m\rangle$, and $\left|D_{\mathrm{rs}}\right\rangle$ on the equators with the azimuthal angles $2 \alpha_{\mathrm{p}}, 2 \alpha_{\mathrm{m}}$, and $2 \alpha_{\mathrm{rs}}$, respectively. Here, $\alpha_{\mathrm{p}}\left(\alpha_{\mathrm{m}}\right)$ is the orientation angle of the electric (magnetic) dipole in the coordinate frame; $\alpha_{\mathrm{rs}}$ is the orientation angle of the RS dipole in the constitutive frame. The meta-atoms in Fig. \ref{fig4}\textbf{a} are designed to satisfy $\alpha_{\mathrm{p}}=\alpha_{\mathrm{rs}}$. The metasurface generates the scattering field $\boldsymbol{\Psi}_{\text {out }}=[(\hat{\mathbf{e}}_x+\mathrm{i} \sigma_{\text {out}} \hat{\mathbf{e}}_y)+\mathrm{i} \kappa_{\text {out}}(\hat{\mathbf{h}}_y-\mathrm{i} \sigma_{\text {out}} \hat{\mathbf{h}}_x)] E e^{\mathrm{i} \kappa_{\text {out}} k_0 z-\mathrm{i} \omega t}$, which has four polarization components $\left|\sigma_{\text {out}}= \pm 1, \kappa_{\text {out}}= \pm 1\right\rangle$, as shown in Fig. \ref{fig4}\textbf{a}. Correspondingly, there are four polarization evolution channels $\left|\sigma_{\mathrm{in}}, \kappa_{\mathrm{in}}\right\rangle \rightarrow \left|\sigma_{\text {out}}, \kappa_{\text {out}}\right\rangle$. Figure \ref{fig4}\textbf{b} shows the 4D polarization evolution in the channel $|-1,-1\rangle \rightarrow|+1,+1\rangle$ as an example. The resulting geometric phase is:
\begin{equation}
\Phi_{\mathrm{tot}}=\Phi_{\mathrm{pb}}+\Phi_{\mathrm{rs}}=\left(\sigma_{\mathrm{in}}-\sigma_{\mathrm{out}}\right) \alpha_{\mathrm{p}}+\left(\kappa_{\mathrm{in}}-\kappa_{\mathrm{out}}\right) \alpha_{\mathrm{rs}}.
     \label{eq:1}
\end{equation}
Figure \ref{fig4}\textbf{c} shows the phases \textcolor{black}{and amplitudes} of the four output waves for different orientation angle $\alpha_{\mathrm{rs}}$. The simulated phases (solid lines) are consistent with the total geometric phases (symbols) given by the solid angles on the Poincaré hypersphere, \textcolor{black}{confirming the emergence of both the PB and RS phases.} The simulated amplitudes of different outputs are approximately equal and nearly independent of $\alpha_{\mathrm{rs}}$. 

Figure \ref{fig4}\textbf{d} shows the simulated electric fields of the four output waves. The wavefronts are consistent with the deflection directions (yellow arrows) predicted based on the geometric phase gradient. Specifically, the output $|+1,+1\rangle$ exhibits the largest deflection angle; $|+1,-1\rangle$ and
$|-1,+1\rangle$ exhibit the same deflection angle; $|-1,-1\rangle$ undergoes no deflection. The number of distinct output wavefronts can be increased by flipping the electric spin and RS spin of the incident wave (Supplementary Notes VII and VIII). Figure \ref{fig4}\textbf{e} shows the fabricated metasurface sample comprising two supercells. \textcolor{black}{The scattering far-field intensities under the incidence of the plane waves $\left|\sigma_{\mathrm{in}}= \pm 1, \kappa_{\mathrm{in}}=-1\right\rangle$ are shown in Fig. \ref{fig4}\textbf{f}}. We notice that a total of eight output intensity lobes (four lobes in each case) emerge in six distinct directions, with consistency between the experimental (red lines) and simulation (black lines) results. Unlike the case in Fig. \ref{fig4}\textbf{c}, the intensities of different outputs are not equal due to the finite size of the metasurface.

\textcolor{black}{The output wavefronts can be further increased in the higher-order RS metasurface with $\alpha_{\mathrm{p}}=\alpha_{\mathrm{m}} \neq \alpha_{\mathrm{rs}}$.} Under the incidence of the plane waves $\left|\sigma_{\mathrm{in}}= \pm 1, \kappa_{\mathrm{in}}= \pm 1\right\rangle$, this metasurface can generate twelve output wavefronts propagating in different directions, as schematically shown in Fig. \ref{fig5}\textbf{a}. The simulated far-field intensity patterns are shown in Fig. \ref{fig5}\textbf{b}, where the intensity lobes are labelled in accordance with the numbers in Fig. \ref{fig5}\textbf{a}. As seen, the scattered far fields can propagate in twelve different directions, depending on the values of $\sigma_{\text {out}}$ and $\kappa_{\text {out}}$. 
Additionally, the number of distinct outputs can be dynamically reconfigured by employing the incident wave  $\left|\Psi_{\text {in}}\right\rangle=\eta_1|+1,+1\rangle+\eta_2|+1,-1\rangle+\eta_3|-1,+1\rangle+\eta_4|-1,-1\rangle$ with varying 4D polarization. For example, by tuning the coefficients $\eta_{1-4}$ to obtain four different incident polarizations, denoted by the points “i-iv” on the Poincaré hypersphere in Fig. \ref{fig5}\textbf{c}, the metasurface can generate 12, 10, 8, and 6 far-field intensity lobes, as shown in Fig. \ref{fig5}\textbf{d}. These results demonstrate the practical application potential of the framework, \textcolor{black}{which can be further enriched by introducing time-varying or active components into the system} \cite{zhang2018space,shaltout2019spatiotemporal,galiffi2022photonics,tirole2023double}.

\section*{Discussion}
\noindent \textcolor{black}{We unveil the geometric phases arising from complete electromagnetic polarization evolution in the 4D RS space by treating electromagnetic waves as a four-component bispinor. Beyond the conventional PB phase induced by electric or magnetic polarization evolution, we discover a new geometric phase, the RS phase, which originates from the hybrid electric-magnetic polarization evolution and can manifest even in linearly polarized waves. The complementary RS and PB phases enable a unified high-dimensional geometric framework for understanding and controlling phase shifts in light propagation across general interfaces, including artificial metasurfaces. The proposed mechanism applies to general electromagnetic waves, including evanescent waves and complex structured waves. Our work broadens the geometric phase paradigm and offers fundamental insights into the geometric nature of light-matter interactions, opening new avenues for exploring topological and non-Abelian phenomena in high-dimensional classical wave systems. Future research may integrate the spin-redirection geometric phase \cite{chiao1986manifestations,RN45} arising from variations of propagation direction into this framework, which promises a more comprehensive approach for geometric phase physics.}

\textcolor{black}{After the initial posting of this work \cite{cheng2025riemann}, a related preprint appeared \cite{vernon2025electric}, proposing an “electric-magnetic (EM) geometric phase” arising from cyclic evolution of the $\mathbf{E}$--$\mathbf{H}$ relationship in nonparaxial light. This EM geometric phase is conceptually equivalent to the RS geometric phase described here and can be regarded as a special case within our broader high-dimensional framework based on the 4D RS space. The RS geometric phase applies to general electromagnetic waves, including nonparaxial waves (e.g., evanescent waves and structured waves). Their conclusion that EM geometric phase appears exclusively in nonparaxial light does not conflict with our results, as their analysis does not address the scenarios involving material interfaces, where the local $\mathbf{E}$--$\mathbf{H}$ relationship can be modified. We view their work as complementary to ours, together contributing to a more complete understanding of this geometric phase across different wave regimes.}

\section*{Methods}
\subsection*{Poincaré hypersphere representation}\ \\ 
The 4D electromagnetic polarization can be represented on the Poincaré hypersphere in Fig. \ref{fig1}\textbf{a}. The north and south poles of the E-sphere are $\left|N_{\mathrm{e}}\right\rangle=\frac{1}{\sqrt{2}}\left(\hat{\mathbf{e}}_i+\mathrm{i} \hat{\mathbf{e}}_j\right)$ and $\left|S_{\mathrm{e}}\right\rangle=\frac{1}{\sqrt{2}}\left(\hat{\mathbf{e}}_i-\mathrm{i} \hat{\mathbf{e}}_j\right)$, respectively. The north and south poles of the H-sphere are $\left|N_{\mathrm{m}}\right\rangle=\frac{1}{\sqrt{2}}(\hat{\mathbf{h}}_j-\mathrm{i} \hat{\mathbf{h}}_i)$ and $\left|S_{\mathrm{m}}\right\rangle= \frac{1}{\sqrt{2}}(\hat{\mathbf{h}}_j+\mathrm{i} \hat{\mathbf{h}}_i)$, respectively. Any 4D polarization state $|\Psi\rangle$ can be parameterized by six parameters ($\theta_{\mathrm{e}}, \phi_{\mathrm{e}}, \theta_{\mathrm{m}}, \phi_{\mathrm{m}}, \theta_{\mathrm{rs}}, \phi_{\mathrm{rs}}$) as

\begin{equation}
\begin{aligned}
     |\Psi\rangle & =\frac{1}{\sqrt{2}}\left[\cos \left(\frac{\theta_{\mathrm{rs}}}{2}\right) e^{-\mathrm{i} \frac{\phi_{\mathrm{rs}}}{2}}+\sin \left(\frac{\theta_{\mathrm{rs}}}{2}\right) e^{\mathrm{i} \frac{\phi_{\mathrm{rs}}}{2}}\right]\left|H_{\mathrm{rs}}\right\rangle \\
    & +\mathrm{i} \frac{1}{\sqrt{2}}\left[\cos \left(\frac{\theta_{\mathrm{rs}}}{2}\right) e^{-\mathrm{i} \frac{\phi_{\mathrm{rs}}}{2}}-\sin \left(\frac{\theta_{\mathrm{rs}}}{2}\right) e^{\mathrm{i} \frac{\phi_{\mathrm{rs}}}{2}}\right]\left|V_{\mathrm{rs}}\right\rangle,
     \label{eq:2}
\end{aligned}
\end{equation}
where $\left|H_{\mathrm{rs}}\right\rangle=\cos \left(\frac{\theta_{\mathrm{e}}}{2}\right) e^{-\frac{\mathrm{i} \phi_{\mathrm{e}}}{2}}\left|N_{\mathrm{e}}\right\rangle+\sin \left(\frac{\theta_{\mathrm{e}}}{2}\right) e^{\frac{\mathrm{i} \phi_{\mathrm{e}}}{2}}\left|S_{\mathrm{e}}\right\rangle$ represents an arbitrary state on the E-sphere and $\left|V_{\mathrm{rs}}\right\rangle=\cos \left(\frac{\theta_{\mathrm{m}}}{2}\right) e^{-\frac{\mathrm{i} \phi_{\mathrm{m}}}{2}}\left|N_{\mathrm{m}}\right\rangle+\sin \left(\frac{\theta_{\mathrm{m}}}{2}\right) e^{\frac{\mathrm{i} \phi_{\mathrm{m}}}{2}}\left|S_{\mathrm{m}}\right\rangle$ represents an arbitrary state on the H-sphere. The representation of the electric (magnetic) polarization on the E-sphere (H-sphere) has been well established in the literature. The representation of the hybrid $\mathbf{E}$-$\mathbf{H}$ polarization on the RS-sphere can be understood by considering the RS vector $\mathbf{F}_1=E_i \hat{\mathbf{e}}_i+\mathrm{i} H_j \hat{\mathbf{h}}_j$ of paraxial waves as an example ($\mathbf{F}_2$ exhibits the same polarization and can be represented similarly). The temporal evolution of $\mathbf{F}_1$ traces out a polarization ellipse on the plane with the bases ($\hat{\mathbf{e}}_i, \hat{\mathbf{h}}_j$), where the polarization ellipticity and orientation are determined by the relative magnitude and phase of $E_i$ and $H_j$. The polarization of $\mathbf{F}_1$ can be mapped to a point on the RS-sphere with the normalized RS Stokes vector $\left(S_1, S_2, S_3\right)$, where $S_1=\frac{\left|E_i\right|^2-\left|H_j\right|^2}{\left|E_i\right|^2+\left|H_j\right|^2}, S_2=\frac{2 \operatorname{Re}\left[E_i\left(\mathrm{i} H_j\right)^*\right]}{\left|E_i\right|^2+\left|H_j\right|^2}=\frac{2 \operatorname{Im}\left[E_i H_j^*\right]}{\left|E_i\right|^2+\left|H_j\right|^2}$, and $S_3= \frac{-2 \operatorname{Im}\left[E_i\left(\mathrm{i} H_j\right)^*\right]}{\left|E_i\right|^2+\left|H_j\right|^2}=\frac{2 \operatorname{Re}\left[E_i H_j^*\right]}{\left|E_i\right|^2+\left|H_j\right|^2}$. Importantly, $S_1$ is determined by the relative amplitude of $E_i$ and $H_j ; S_2$ and $S_3$ are determined by the imaginary and real parts of the complex Poynting vector, respectively.

Some representative RS polarization states and corresponding $E_i$-$H_j$ relations are shown in Fig.
S1. The points located at $\phi=0$ longitudinal line correspond to propagating waves in transparent media with a real Poynting vector. Specifically, the north and south poles denote the free-space propagating waves in opposite directions. The points on the equator represent purely evanescent waves with an imaginary Poynting vector, e.g., waves in lossless electric/magnetic plasma media. The remaining points on the sphere denote waves in lossy or gain media with a complex Poynting vector.

\subsection*{Numerical simulation}\

\noindent All the full-wave numerical simulations are performed with the package COMSOL Multiphysics. In the simulation of the metasurfaces in Figs. \ref{fig3}-\ref{fig5}, we set the periods of unit cells along $x$ and $y$ directions to be $L_x=L_y=5.5 \mathrm{~mm}$. The working frequency is 24 GHz. In each unit cell, the split ring has two arms with geometric parameters: $t_1=0.035 \mathrm{~mm}, t_2=0.15 \mathrm{~mm}, h=1 \mathrm{~mm}, w=2.3 \mathrm{~mm}, l=0.58 \mathrm{~mm}$, and $g=0.27 \mathrm{~mm}$ (refer to Fig. \ref{fig3}\textbf{a} for the definition of the geometric parameters). The split rings are made of copper with the electrical conductivity $\sigma= 5.814 \times 10^7 \mathrm{~S} / \mathrm{m}$. To change the 4D dipole polarization of the meta-atom, we rotate the split ring around the $x, y$, and $z$ axis by angles $\alpha_x, \alpha_y$, and $\alpha_z$, respectively. For the numerical demonstration in Fig. \ref{fig3}, each supercell comprises six split rings with the rotation angles $(\alpha_x, \alpha_y, \alpha_z)$ in degrees: $(0,-90,0)$, $(0,-68,0)$, $(0,0,0)$, $(-90,0,0)$, $(0,180,0)$, and $(0,112,0)$. For the numerical demonstration in Fig. \ref{fig4}, each supercell comprises six split rings with the rotation angles $(\alpha_x, \alpha_y, \alpha_z)$ in degrees: $(0,-90,-90)$, $(0,-68,-60)$, $(0,0,-30)$, $(-90,0,0)$, $(0,180,30)$, and $(0,112,60)$. For the numerical demonstration in Fig. \ref{fig5}, each supercell comprises six split rings with the rotation angles $(\alpha_x, \alpha_y, \alpha_z)$ in degrees: $(0,82,-160)$, $(0,68,-120)$, $(0,45,-80)$, $(-45,0,-40)$, $(-90,0,0)$, $(-130,0,40)$, $(0,135,80)$, $(0,112,120)$, and $(0,96,160)$. To obtain the incident 4D polarizations “i-iv” in Fig. \ref{fig5}\textbf{c}, we set the coefficients ($\eta_1$, $\eta_2$, $\eta_3$, $\eta_4$)=($\frac{1}{2}$,$\frac{1}{2}$,$\frac{1}{2}$,$\frac{1}{2}$) for “i”, ($\frac{1}{\sqrt{3}}$,$\frac{1}{\sqrt{3}}$,$\frac{1}{\sqrt{3}}$,0) for “ii”, (0,$\frac{1}{\sqrt{2}}$,$\frac{1}{\sqrt{2}}$,0) for “iii”, and ($\frac{1}{\sqrt{2}}$,0,$\frac{1}{\sqrt{2}}$,0) for “iv”.

\subsection*{Experiment}\ 

\noindent The metasurfaces are fabricated on a Roger's 5880 substrate with printed circuit board technology (thickness $t_{\mathrm{s}}=0.508 \mathrm{~mm}$, height $h_{\mathrm{s}}=4 \mathrm{~mm}$, relative permittivity $\varepsilon_r=2.2$, and loss $\operatorname{tangent} \tan \delta=0.0009$). Experimental characterization is performed in a microwave anechoic chamber to suppress multi-path effects. The setup comprises a linearly polarized transmitting horn antenna, a receiving horn antenna (Rx), and a vector network analyzer (VNA, Keysight PNA 5227B). Both antennas are positioned 1 m from the metasurface and connected to the two ports of the VNA via $50 \Omega$ coaxial cables. By rotating the angular position of the Rx horn antenna with respect to the metasurface and measuring the transmitted/reflected signals by the VNA, the far-field scattering pattern of the metasurface is obtained. For the measurements in Fig. \ref{fig4}, wideband 3D-printed polarizers are mounted on the horn apertures to generate and detect circularly polarized waves. Prior to characterizing the metasurface, a background signal measurement is performed without the sample to capture the direct coupling between the source and receiver as well as other ambient contributions. The measured background signals are then used for calibration to minimize the contributions of these spurious signals to the measurement.

\section*{Acknowledgements}
\noindent The work described in this paper was supported by grants from the Research Grants Council of the Hong Kong Special Administrative Region, China (Projects Nos. AoE/P-502/20 and CityU11308223) and National Natural Science Foundation of China (No. 12322416). D.P.T. acknowledges support from the Research Grants Council of the Hong Kong Special Administrative Region, China (Project Nos. C5031-22G, C5078-24G, CityU11305223, CityU11300224, CityU11304925, and CityU11305125), City University of Hong Kong (Project No. 9380131), and National Natural Science Foundation of China (Grant No. 62375232). G.-B.W. acknowledges support from the Research Grants Council of the Hong Kong Special Administrative Region, China (Project No. CityU21207824). \textcolor{black}{The authors thank Prof. C. T. Chan and Prof. Z. Q. Zhang for helpful discussions.} The authors also thank Dr. Ka Fai Chan and Dr. Chenfeng Yang for their support in the experiments. 

\section*{Author contributions}
\noindent S.W. and Y.C. conceived the idea and developed the concepts. Y.C. designed the structures and conducted the numerical simulations. Y.-S.Z., Y.C., and G.-B.W. designed and carried out the experiments. Y.C. and S.W. wrote the draft. S.W., G.-B.W., and D.P.T. supervised the project. All authors contributed to discussions\textcolor{black}{, interpretation of the results, and polishing of the manuscript}.

\renewcommand\bibnumfmt[1]{#1.}
\def\bibsection{}
\section*{References}
\bibliography{MyCollection}
\bibliographystyle{reference}

\end{document}